\documentclass[12pt]{article}

\usepackage{latexsym,amsmath,amssymb,amscd,eucal}
\usepackage{xy}
\usepackage{enumerate}

\input xy
\xyoption{all}

\def\harr#1#2{\smash{\mathop{\hbox to .3in{\rightarrowfill}}
 \limits^{\scriptstyle#1}_{\scriptstyle#2}}}

\def\appendix#1{\addtocounter{section}{1}\setcounter{equation}{0}
\renewcommand{\thesection}{\Alph{section}}
\section*{Appendix \thesection\protect\indent \parbox[t]{11.715cm} {#1}}
\addcontentsline{toc}{section}{Appendix \thesection\ \ \ #1} }
\newcommand{\eq}{\begin{equation}}
\newcommand{\eqend}{\end{equation}}

\newbox\ncintdbox \newbox\ncinttbox
\setbox0=\hbox{$-$} \setbox2=\hbox{$\displaystyle\int$}
\setbox\ncintdbox=\hbox{\rlap{\hbox to
\wd2{\hskip-.125em\box2\relax \hfil}}\box0\kern.1em}
\setbox0=\hbox{$\vcenter{\hrule width 4pt}$}
\setbox2=\hbox{$\textstyle\int$}
\setbox\ncinttbox=\hbox{\rlap{\hbox to
\wd2{\hskip-.175em\box2\relax \hfil}}\box0\kern.1em}

\def\be{\begin{equation}}
\def\ee{\end{equation}}
\def\bea{\begin{eqnarray}}
\def\eea{\end{eqnarray}}
\def\bd{\begin{displaymath}}
\def\ed{\end{displaymath}}

\DeclareFontFamily{U}{rsf}{}
\DeclareFontShape{U}{rsf}{m}{n}{
  <5> <6> rsfs5 <7> <8> <9> rsfs7 <10-> rsfs10}{}
\DeclareMathAlphabet\Scr{U}{rsf}{m}{n}

\def\cP{{\Scr P}}

\def\cE{{\Scr E}}
\def\cF{{\Scr F}}

\def\cO{{\Scr O}}

\makeatletter
\newdimen\normalarrayskip              
\newdimen\minarrayskip                 
\normalarrayskip\baselineskip
\minarrayskip\jot
\newif\ifold             \oldtrue            
\def\arraymode{\ifold\relax\else\displaystyle\fi} 
\def\@arrayskip{\ifold\baselineskip\z@\lineskip\z@
     \else
     \baselineskip\minarrayskip\lineskip2\minarrayskip\fi}
\def\@arrayclassz{\ifcase \@lastchclass \@acolampacol \or
\@ampacol \or \or \or \@addamp \or
   \@acolampacol \or \@firstampfalse \@acol \fi
\edef\@preamble{\@preamble
  \ifcase \@chnum
     \hfil$\relax\arraymode\@sharp$\hfil
     \or $\relax\arraymode\@sharp$\hfil
     \or \hfil$\relax\arraymode\@sharp$\fi}}
\def\@array[#1]#2{\setbox\@arstrutbox=\hbox{\vrule
     height\arraystretch \ht\strutbox
     depth\arraystretch \dp\strutbox
     width\z@}\@mkpream{#2}\edef\@preamble{\halign \noexpand\@halignto
\bgroup \tabskip\z@ \@arstrut \@preamble \tabskip\z@ \cr}%
\let\@startpbox\@@startpbox \let\@endpbox\@@endpbox
  \if #1t\vtop \else \if#1b\vbox \else \vcenter \fi\fi
  \bgroup \let\par\relax
  \let\@sharp##\let\protect\relax
  \@arrayskip\@preamble}
\makeatother

\newcommand{\beq}{\begin{eqnarray}}
\newcommand{\eeq}{\end{eqnarray}}

\def\appendix#1{\addtocounter{section}{1}\setcounter{equation}{0}
\renewcommand{\thesection}{\Alph{section}}
\section*{Appendix \thesection. #1}
\addcontentsline{toc}{section}{Appendix \thesection\ \ \ #1} }

\newtheorem{remark}{Remark}[section]

\numberwithin{equation}{section}

\begin{document}

\begin{center}

{\Large\bf BRST-Invariant Deformations of Geometric Structures in Sigma Models}

\end{center}
\vspace{0.1in}

\begin{center}
{\large
A. A. Bytsenko 
\footnote{abyts@uel.br}


\vspace{10mm}
{\it Departamento de F\'{i}sica, Universidade Estadual deLondrina, Caixa Postal 6001, Londrina-PR, Brazil}
}


\end{center}

\vspace{0.1in}
\begin{center}
{\bf Abstract}
\end{center}
We study a Lie algebra of formal vector fields $W_n$ with its application to the perturbative deformed holomorphic symplectic structure in the A-model, and a Calabi-Yau manifold with boundaries in the B-model. We show that equivalent classes of deformations are describing by a Hochschild cohomology theory of the DG-algebra 
${\mathfrak A} = (A, Q)$, $Q =\overline{\partial}+\partial_{\rm deform}$, which is defined to be the cohomology of $(-1)^n Q +d_{\rm Hoch}$. Here
$\overline{\partial}$ is the initial non-deformed BRST operator
while $\partial_{\rm deform}$ is the deformed part whose algebra is a Lie algebra of linear vector fields ${\rm gl}_n$. We show that
equivalent classes of deformations are described by a Hochschild cohomology of ${\mathfrak A}$, an important geometric invariant of the (anti)holomorphic structure on $X$. We discuss the identification of the harmonic structure $(HT^\bullet(X); H\Omega_\bullet(X))$ of affine space $X$ and the group ${\rm Ext}_{X²}^n({\cO}_{\triangle}, 
{\cO}_{\triangle})$ (the HKR isomorphism), and bulk-boundary deformation pairing.



\newpage

\section{Introduction}

Two-dimensional topological field theories have been actively studied (specially in connection with mirror symmetry), however these theories is much more nontrivial and more interesting when can be defined for deformed holomorphic symplectic structure of the cotangent bundle (the A-model) and a Calabi-Yau manifold with boundaries (the B-model).
Recall that original topological field model in arbitrary dimension has been analyzed in \cite{Atiyah} and this theory did not allow for boundaries (but it allow for defects of higher codimension, Wilson loops in Chern-Simons theory, for example). Nontrivial variety of boundary conditions which could be associated with topological branes in the B-model has been introduced in \cite{Witten}.

In this paper we study (perturbative) deformations of complex structure for topological sigma models. A brief summary of the results of the paper is the following.
We study a Lie algebra of formal vector fields $W_n$ with it application to the deformed BRST operator. We consider infinite-dimensional Lie algebra
$W_n\ltimes {\overline G}\otimes P_n$, i.e. a semidirect sum of algebra $W_n$ extended by product of Lie algebra $\overline G$ 
(related to the initial non-deformed BRST operator $\overline{\partial}$) and formal power series $P_n$ of $n$ compex variables $\{\overline{z}_j\}_{j=1}^n$.
Lie algebra $W_n\ltimes {\overline G}\otimes P_n$ contains a small Lie subalgebra ${\rm g}{\rm l}_n\oplus {\overline G}$, which consists of linear {\it vector fields} and ${\overline G}$-valued fields. We show that on the bigraded vector space $C^\bullet(A)$ the Hochschild cohomology of an associative algebra $A$ is defined to be the cohomology of $(-1)^n (\overline{\partial} + \partial_{\rm deform})+d_{\rm Hoch}$ (Lie algebra of the operator $\partial_{\rm deform}$ is ${\rm gl}_n$). We also discuss the identification of the harmonic structure $(HT^\bullet(X); H\Omega_\bullet(X))$ of (affine) space $X$ and the group ${\rm Ext}_{X²}^n({\cO}_{\triangle}, {\cO}_{\triangle})$ (the Hochschild-Kostant-Rosenberg isomorphism (HKR)) and bulk-boundary deformation pairing.

\section{Two-categories of sigma-models and their deformations}
\label{Two-categories}
In this section we will mostly followed \cite{Kapustin08,Kapustin09} in reproducing of necessary results.
Let ${{\mathcal X}} =(X,s)$ be a pair in which $X$ is a real manifold and $s$ is a geometric structure on $X$ (such as a complex structure or a symplectic structure). Let $\Sigma$ be a real $N$-dimensional manifold. A topological sigma-model with world-volume $\Sigma$ and a target space ${\mathcal X}$ is a quantum field theory based on a path integral (the measure on the space of maps $\Sigma \rightarrow X$ is dertermined by the structure $s$). ${\mathcal X}$ with a certain type structure serve as target-spaces for topological sigma-model form a $N$-category $\mathfrak C$ with proper features.

Recall that the category ${\mathfrak C}$, associated with pair $(X, s)$, has a symmetric monoidal structure related to the cartesian product of manifolds:
$
{\mathfrak C}\times{\mathfrak C}\rightarrow{\mathfrak C},\,
(X_1, s_1)\times(X_2,s_2) = (X_1\times X_2, s_1\times s_2),
$
with the natural structure $s_1\times s_2$  on $X_1\times X_2$.
This monoidal structure has a unit element ${\mathcal X}_{pt}=(X_{pt}, s_{pt})$, where $s_{pt}$
is the corresponding trivial structure of the manifold consisting of a single point $X_{pt}$. It is clear that
${\mathcal X}_{pt}\times {\mathcal X}= {\mathcal X}$. Define a $(N-1)$-category of morphisms 
${\mathfrak C} := {\rm Hom}_{{\mathfrak C}}({\mathcal X}_{pt}, {\mathcal X})$.
In quantum field theory this categogy is known as the category of boundary conditions of the topological sigma model associated with ${\mathcal X}$. The category ${\mathfrak C}$ has a contravariant
duality functor
$
{\mathfrak C} \stackrel{\lozenge}{\rightarrow} {\mathfrak C},\,
(X,s)^{\lozenge} = (X, s^{\lozenge}),
$
such that there is a canonical equivalence between $(N-1)$-categories
of morphisms:
$
{\rm Hom}_{{\mathfrak C}}({\mathcal X}_1,{\mathcal X}_2) = {\rm Hom}_ {{\mathfrak C}} ({\mathcal X}_{pt}, {\mathcal X}^{\lozenge}_1\times {\mathcal X}_2)
={\mathfrak C}_{{\mathcal X}_1^{\lozenge}\times {\mathcal X}_2}.
$
This equivalence implies that an object
${\mathfrak E}_{12}\in {\rm Hom}_{{\mathfrak C}}({\mathcal X}_1,{\mathcal X}_2)$ determines a functor between $(N-1)$-categories
$
F [{\mathfrak E}_{12}]: \,{\mathfrak C}_{{\mathcal X}_1} \rightarrow {\mathfrak C}_{{\mathcal X}_2},
$
which represents a composition of morphisms within ${\mathfrak C}$.
Moreover, a composition of morphisms of ${\mathfrak C}$ corresponds to the composition of functors, so the structure of the $N$-category ${\mathfrak C}$ is
determined by the boundary condition categories ${\mathfrak C}_{{\mathcal X}}$ and the
functors.

\subsection{The A-model and deformations of holomorphic symplectic structures}

Let ${\mathfrak C}$ be a $(N=2)$-category. For the A-model the structure $s$ is a symplectic structure (that is, $s$ is a symplectic form on $X$), the category of boundary conditions ${\mathfrak C}_{{\mathcal X}}$ is the Fukaya-Floer category, its simplest objects being lagrangian submanifolds of $X$, the action of the duality functor is $s^{\lozenge}=- s$, and the functor $F [{\mathfrak E}_{12}]$ is the lagrangian correspondence functor determined by a lagrangian submanifold $M_{12}\subset{\mathcal X}_1^{\lozenge}\times {\mathcal X}_2$.
Deformations of holomorphic symplectic structure which preserve the de Rham cohomology class of $\omega$ 
are parameterized up to gauge equivalence by Maurer-Cartan elements of the differential Poisson algebra ${\cP} (X, \omega)$. We refer the reader to the papers \cite{Kapustin08,Kapustin09}, where the procedure of perturbative deformations for the A-model has been analyzed in details. In the next section we briefly discuss the descent procedure of deformation in the B-model.

\subsection{The B-Model and the descent procedure of deformation}

For the B-model $X$ is a Calabi-Yau manifold, $s$ is its complex structure, the category of boundary conditions ${\mathfrak C}_{{\mathcal X}}$ is the bounded derived category of coherent sheaves
${\rm D}^b_{\rm coh}({\mathcal X})$, its simplest objects being complexes of holomorphic
vector bundles on $X$. The duality functor acts trivially:
$s^\lozenge =s$, and the functor $F[{\mathfrak E}_{12}]$ is the
Fourier-Mukai transform corresponding to the object ${\mathfrak E}_{12}$.
Note that all suitable enough functors between the derived category of quasicoherent sheaves on two varieties, ${\mathcal X}_1$ and ${\mathcal X}_2$, are given by Fourier-Mukai transforms. 
\\

\noindent
{\bf Local topological observables.}
Recall that for the topological field theory the transformation law in terms of BRST operator $Q$ can be expressed as $\delta W = -\{Q, W\}$
for any field $W$; in addition $Q^2 = 0$. Standard arguments using the $Q$ invariance show that the correlation function $\langle \{Q, W\}\rangle = 0$ for any $W$. It is also true that $\{Q, {{\cO}}_a\} = 0$ for some BRST invariant operators ${{\cO}}_a$. The local variables are
$
\eta^{\bar i} \in {\Gamma}(\phi^*T^{\vee}X) 
\sim d{\bar z}^{\bar i},
$
$
\theta_i \in g_{i{\bar j}}({\Gamma}(\phi^*T^{\vee}X))^{\bar j}
\sim \partial/{\partial z^i}\,
$
(see \cite{Witten,Sharpe03} for more details).
Mathematically one can interpret $\eta^{\overline{i}}$ as the (0, 1) form $d\phi^{\overline{i}}$, and $\phi^{\overline{i}} \sim \overline{z}^{\overline{i}},\,
\eta^{\overline{i}}\sim d\phi^{\overline{i}}$.
The Noeter BRST charge $Q$ acts on the fields as 
$
Q\sim \overline{\partial}  =  \sum_i\eta^{\overline i}\partial/\partial\phi^{\overline i}
\sim \sum_id{\overline z}^{\overline i}
\partial/\partial {\overline {z}}^{\overline i}.
$
The sheaf cohomology group $H^p(X,\Lambda^qTX)$ associated with $(0,q)$ forms on space $X$ with values in $\Lambda^qTX$ (the $q^{th}$ exterior power of the holomorphic tangent bundle of $X$),
consists of solutions of $\overline{\partial}V=0$ modulo $V\rightarrow V + \overline{\partial} S$ for an object
\begin{eqnarray}
{{{\cO}}}_V & = & \eta^{\bar i_1}\eta^{\bar i_2}\dots \eta^{\bar i_p}V_{{\bar i_1}{\bar i_2}\dots
\bar i_p}{}^{j_1 j_2\dots j_q}\psi_{j_1}\dots \psi_{j_q}
\nonumber \\
& = &
d\bar z^{{\bar i}_1}d\bar z^{{\bar i}_2}\dots d\bar z^{{\bar i}_p}V_{\bar i_1\,\bar i_2
\dots \bar i_p}{}^{j_1j_2\dots j_q}{\partial/\partial z_{j_1}}
\dots {\partial/\partial z_{j_q}}\,,
\label{operator}
\end{eqnarray}
which is called the vertex operator, and some $S$.
As a consequence,
\begin{equation}
\delta_Q({\cO}_V) = {{\cO}}_{\overline{\partial} V};\,\,\,\,\, 
{{\cO}}_V\,\,\, {\rm is}\,\,\, {\rm BRST}\,\,\, 
{\rm invariant}\,\,\, {\rm if}\,\,\, {\overline{\partial}}V=0
\,\,\, {\rm and}\,\,\, {\rm BRST}\,\,\, {\rm exact}\,\,\, 
{\rm if}\,\,\, V=\overline{\partial} S\,.
\label{vertex}
\end{equation}
The BRST transformation (for bosonic and fermionic fields) is nilpotent, $\delta_Q^2 = 0$, and is a derivation of the algebra generated by fields and their derivatives. There is a natural map
$V\rightarrow {{\cO}}_V$ from 
$\oplus_{p,q}H^p(X,\Lambda^q TX)$ to the BRST cohomology of the B-model. Therefore the BRST cohomology is isomorphic to the Dolbeault cohomology, and {\it on the classical level} the BRST operator acts as Dolbeault operator.
\\
\\
\noindent
{\bf The descent procedure.}
A systematic way to construct deformations of boundary conditions is the descent procedure \cite{Witten}. If ${\cO}$ is an even (bosonic) local observable on the boundary then its descendants ${{\cO}}^{(1)}\in \Omega^1(X)$ and ${{\cO}}^{(2)}\in \Omega^2(X)$ are defined by 
\begin{equation}
d {\cO} = \delta_Q {\cO}^{(1)} + \ldots , \,\,\,\,\,\,\,\,\,\,\,
d {\cO}^{(1)} = \delta_Q {\cO}^{(2)} + \ldots 
\end{equation}
Then in forming the topological family the generalized action becomes
$
S_{\rm bulk} \Longrightarrow S_{\rm bulk} + \gamma \int_{\Sigma}{{\cO}}^{(2)},
$
where $\gamma$ is a formal parameter. Because of the definition of ${{\cO}}^{(2)}$ it follows that the modified action is BRST-invariant up to terms proportional to equations of motion. It is clear that local topological observables on the boundary are of the same $(0, q)$-form as in the bulk. 
An element of space of infinitisemal deformations can be represented by a $\overline{\partial}$-closed inhomogeneous form $W$ of even degree. Thus the corresponding observable can be thought of as an even function $W(\phi, \eta)$ of bosonic variables $\phi^i,\,  \phi^{\overline{i}}$ and fermionic variables $\eta^i$, it satisfying 
\begin{equation}
Q\,W = \sum_i{\eta}^{\overline i}
\frac{\partial W(\phi, \eta)}{\partial {\phi}^{\overline i}} 
= 0\,.
\end{equation}
Next the descendans of $W$ have to be constructed. Note that {\it on the quantum level} one requires $X$ to be a Calabi-Yau manifold, i.e. one requires the existence of a holomorphic volume form on $X$.
The volume form is used to write down a BRST-invariant measure on the space of bosonic and fermionic zero modes.

Let $W$ be a holomorphic function on a manifold $X$. One can deform the B-model with target $X$ by modifying the transformation law for fermionic fields and adding to the action a new term which is related to this deformation (see for detail \cite{Kapustin08}). This model can be called the Landau-Ginzburg model with target $X$ and superpotential $W$. The algebra of topological observables for the Landau-Ginzburg model is the hypercohomology of the complex
$$
\Lambda^n TX \longrightarrow \Lambda^{n-1} TX \longrightarrow \ldots \longrightarrow TX \longrightarrow {\cO}_X
$$
where the differential is contraction with the holomorphic 1-form
$-\partial W$. Let $X$ be a contractible open subset of
${\mathbb C}^n$ and the critical points of $W$ are isolated, then the
hypercohomology of this complex is the Jacobi algebra
$
H^0(X,{\cO}_X)/A_{\partial W}
$
where $A_{\partial W}$ is the ideal generated by partial derivatives
of $W$. This construction can be generalized by replacing $W$ with an
inhomogeneous even $\overline{\partial}$-closed form on $X$. This generalization is called {\it the curved B-model}. 
It can be thought of as a generalization of the Landau-Ginzburg deformation of the B-model \cite{Kapustin08}. 

Consider a local observable
$W(\phi,\eta)$ representing an even element of
$
\oplus_p H^{p}({\cO}_X).
$
Then one can determed its descendants $W^{(j)}$ and their BRST transformations. More generally, for non-vanish form which is a trivial class in $\overline{\partial}$-cohomology, one can restore BRST-invariance at some order by writing 
$W{(\varepsilon)} = \sum_jW^{(j)}\varepsilon^j$ (where typically $\varepsilon^j$ is of order $\hbar^j$). Finally the total action has to be BRST-invariant. As a result we can get the curved B-model which is ${\mathbb Z}_2$-graded like the Landau-Ginzburg model.
Also the algebra of observables in the curved B-model is computed in
essentially the same way as in the Landau-Ginzburg model. Indeed,
the algebra of observables is the cohomology of a certain
differential $\delta_Q$ in the space of $(0,q)$ forms with values in
polyvector fields of type $(p,0)$. This differential is given by
\begin{equation}
\delta_Q = \overline{\partial}-\partial W\,\llcorner .
\label{deform2}
\end{equation}
If $X$ is compact and K\"ahler, then one can always find a form in the cohomology class of $W$ which is $\partial$-closed. For such $X$ the differential $\delta_Q$ reduces to $\overline{\partial}$, and the algebra of topological observables is the same as in the ordinary B-model.

\section{Lie algebras of formal vector fields}

Suppose that the even function $W(\phi, \eta)$ has the form of formal power series in fields. This situation has been discussed in Section \ref{Two-categories} for the two dimensional sigma-model (perturbative deformations of the holomorphic symplectic structure for the A-model, and the descent procedure of deformations of boundary conditions for the B-model). A systematic way to study deformations is to consider the space of vector fields associate with the deformed BRST operator and its topological algebra.
\\
\\
\noindent
{\bf Lie algebras $W_n^{\overline G}\ltimes {\overline G}\otimes P_n$.}
At more basic level let us consider the infinite-dimensional Lie algebra $W_n^{\overline G}\ltimes {\overline G}\otimes P_n$, i.e.
a semidirect sum of the algebra $W_n^{\overline G}$ extended by the product of the algebra $\overline G$ (related to the operator $\overline{\partial}$) and the formal power series $P_n$ of $n$ variables \cite{Khoroshkin,Bytsenko}. These algebras are interpreted as follows:
\begin{itemize} 
\item{} A Lie algebra $W_n^{\overline G}$ is associated with the deformed part of the BRST operator and it is slightly differ from traditional infinite-dimensional Lie algebra of formal vector fields $W_n$. Its elements in ``coordinates'' can be written in the form $\sum_{i=1}^nF_i \partial/\partial z_i$, where $F_i$ is power series in $z_1, \ldots, z_n$ whose coefficients are elements of the proper algebra $\overline{G}$ of alternating holomorphic forms. Thus when taking the commutator for such ``vector fields'' this fact should be keeped in mind.
\item{} The elements of the algebra $\overline G$ correspond to the initial BRST operator \\
$\overline{\partial} = \sum_{\overline{i}=1}^n d\overline{z}^{\overline{i}}\partial/\partial{\overline{z}^{\overline{i}}}$.
\end{itemize}
A commutator of two elements has the form
\begin{equation}
[v+g_1\otimes p_1, u+g_2 \otimes p_2]  \stackrel{{\rm def}}{=}
[v, u]_{W_n} + [g_1, g_2]_{\overline G}\otimes p_1p_2 + 
g_2\otimes v(p_2) - g_1\otimes u(p_1)\,,
\label{commutator}
\end{equation}
where $v, u \in W_n^{\overline G}$, $g_i\in {\overline G}$, $p_i\in P_n$
and $i\in\{1, 2\}\,$. In the following we consider the cohomology of Lie algebras $W_n$. The corresponding theory is similar to the cohomology of Lie algebras $W_n^{\overline G}$, and the reader will have no difficulty in recovering results for the case of $W_n^{\overline G}$.
The Lie algebra 
$W_n\ltimes {\overline G}\otimes P_n$ contains a Lie subalgebra
which is isomorphic to the direct sum ${\rm g}{\rm l}_n\oplus {\overline G}$. This subalgebra consists of linear vector fields and fixed ${\overline G}$-valued constant fields
\begin{equation}
\iota : \, {\rm g}{\rm l}_n\oplus {\overline G}\hookrightarrow 
W_n\ltimes {\overline G}\otimes P_n \,,\,\,\,\,\,\,\,\,\,\,
\iota (|| a_{i, j}|| + {\overline G})  =  \sum_{i, j}a_{i, j}z_i
\frac{\partial}{\partial z_j} + \sum_{\overline{i}=1}^n d\overline{z}^{\overline{i}}\frac{\partial}
{\partial{\overline{z}^{\overline{i}}}}\otimes {\bf 1}\,.
\label{LAlgebra}
\end{equation}

\section{Morphisms of filtered complexes}

\begin{remark}
{\it Relative cochain complex of the Lie algebra $W_n\ltimes {\overline G}\otimes {P}_n$ with respect to the Lie subalgebra 
${\rm g}{\rm l}_n$ coinsides with the factor-algebra of the relative Weyl algebra over module of the same Lie subalgebra ${\rm g}{\rm l}_n$,
$
{\widetilde W}^\bullet({\rm g}{\rm l}_n\oplus {\overline G}, 
{\rm g}{\rm l}_n) =
W^\bullet({\rm g}{\rm l}_n\oplus {\overline G}, {\rm g}{\rm l}_n)
/F^{2n+1}W^\bullet({\rm g}{\rm l}_n\oplus {\overline G}, 
{\rm g}{\rm l}_n)
$ 
(we refer the reader to paper {\rm \cite{Khoroshkin}} for the proof of this statement). Note that the algebra $\widetilde{W}^\bullet({\rm g}{\rm l}_n\oplus {\overline G})$ has filtration from Weyl algebra {\rm \cite{Fuks}}.} 
\end{remark}
The following main statement asserts the structure of a cohomology ring of the Lie algebra ${\rm g}{\rm l}_n\oplus {\overline G}$:  
The filtered DG-algebra $F^\bullet\widetilde{W}^\bullet({\rm g}{\rm l}_n\oplus {\overline G})$ is quasiconformal to cochain complex of Lie algebra $W_n\ltimes {\overline G}\otimes {P}_n$ with constant coefficients and Serre-Hochshild filtration with respect to Lie subalgebra ${\rm g}{\rm l}_n\oplus {\overline G}$. 
In terms of spectral sequences this statement can be
reformulate as follows. 
Let ${\mathfrak g}\supset {\mathfrak h}$ be a Lie algebra and its subalgebra. Choose expansion ${\mathfrak g}$ into a sum of 
${\mathfrak h}$-modules 
$
{\pi}: \, {\mathfrak g}\stackrel{\sim}{\longrightarrow}
{\mathfrak h}\oplus {\mathfrak g}/{\mathfrak h}\,.
$
Let the same letter ${\pi}: {\mathfrak g}\twoheadrightarrow {\mathfrak h}$ denotes the corresponding projection along ${\mathfrak g}/{\mathfrak h}$. 
By analogy with Leray-Serre filtration \cite{Fuks} one can introduce  filtration $\Phi^\bullet$ on a cochain complex of Lie algebra ${\mathfrak g}$. 
The projection $\pi$ assigns morphism of filtered DG-algebras
\begin{equation}
\overline{\pi}: F^\bullet W^\bullet ({\mathfrak h}) \longrightarrow \Phi^\bullet C^\bullet ({\mathfrak g})\,.
\label{pi1}
\end{equation}
There is a morphism of corresponding spectral sequences and in particular their first terms are:
\begin{equation}
{\pi}^*: H^q({\mathfrak h}; S^p{\mathfrak h}^*)\longrightarrow
H^q({\mathfrak h}; \Lambda^{2p}({\mathfrak g}/{\mathfrak h})^*)\,.
\end{equation}
If Lie algebras ${\mathfrak h}\subset {\mathfrak g}$ have compact Lie groups, then such morphism is a particular case of the Chern-Weyl homomrphism for $H$-fibering 
$G \stackrel{H}{\rightarrow} G/H$, with the following difference: de Rham complexes have to be substituted to cohomological complexes of left invariant forms. Now let us settle an expansion of 
$({\rm g}{\rm l}_n\oplus {\overline G})$-modules
\begin{equation}
\alpha:\, W_n\ltimes {\overline G}\otimes {P}_n \stackrel{\sim}
{\longrightarrow}({\rm g}{\rm l}_n\oplus {\overline G})\oplus
((W_n\ltimes {\overline G}\otimes {P}_n)/
({\rm g}{\rm l}_n\oplus {\overline G}))\,.
\label{alpha}
\end{equation}
Consider 
morphisms of filtered complexes \cite{Khoroshkin}
\begin{eqnarray}
F^\bullet W^\bullet({\rm g}{\rm l}_n\oplus {\overline G})
& \longrightarrow &
F^\bullet {\widetilde W}^\bullet({\rm g}{\rm l}_n
\oplus {\overline G})\,,
\nonumber \\
F^\bullet W^\bullet({\rm g}{\rm l}_n\oplus {\overline G})
& \stackrel{\overline{\alpha}}{\longrightarrow} & 
\Phi^\bullet C^\bullet (W_n\ltimes {\overline G}\otimes {P}_n;
{\bf k})\,,
\label{D}
\end{eqnarray}
here $\overline \alpha$ has been constructed from $\alpha$ as it has been discussed above (formula (\ref{pi1})). The corresponding  morphisms of spectral sequences take the form
\begin{eqnarray}
E^\bullet W^{\bullet, \bullet}({\rm g}{\rm l}_n\oplus 
{\overline G}) 
& \longrightarrow &
E{\widetilde W}^{\bullet, \bullet}({\rm g}{\rm l}_n\oplus 
{\overline G})\,,
\nonumber \\
E^\bullet W^{\bullet, \bullet}({\rm g}{\rm l}_n\oplus 
{\overline G}) 
& \stackrel{\overline{\alpha}}{\longrightarrow} & 
EC^{\bullet, \bullet} (W_n\ltimes {\overline G}\otimes {P}_n;
{\bf k})\,.
\label{D1}
\end{eqnarray}
Spectral sequences in the right hand side of (\ref{D1}) coincide beginning from the first term.

\section{The Hochschild cochain complex and deformations} \label{Hochschild}


Let us consider an algebra $A$ over $\mathbb C$ 
with unit. Let ${\rm gl}_n(A)$ be the Lie algebra of $n\times n$ matrices with coefficients in $A$. 
In fact it is the tensor product $M_n({\mathbb C})\otimes {A}$ of the algebra of $n\times n$ matrices with $A$, considered as a Lie algebra. It contains the Lie subalgebra $M_n({\mathbb C})\otimes {\bf 1}$.
Define the cochain $\Phi^{n} \in C^{k}({\rm gl}_n)$ by the formula 
$
\Phi^n(g_1, \ldots, g_{k}) \stackrel{{\rm def}}{=} 
\sum_{\sigma\in S_{k}}
\mathrm{sign}(\sigma)
\mathrm{Tr}(g_{\sigma(1)} \ldots g_{\sigma(k)}).
$
The homomorphism induced by the standard inclusion 
${\rm gl}_{n-1}\rightarrow {\rm gl}_n$ sends
$\Phi^{n}$ into $\Phi^{n-1}$. The cochain $\Phi^n$ is a 
cocycle \cite{Fuks}. Denote the cohomology class of the cocycle $\Phi^n$
by $\varphi^n$. The tensor product in connection with algebra 
${\rm gl}_n (A)$ becomes $M_n({\mathbb C})\otimes A$. There are chain maps 
\begin{eqnarray}
\varphi^n :\, 
C^\bullet(A) & \longrightarrow &  
C^\bullet({\rm gl}_n (A);\, 
{\rm gl}_n(A)^*),
\\
\varphi^n(\tau)
(M_1\otimes a_1,\ldots, M_k\otimes a_k)
(M_0\otimes a_0)
& = & \sum_{\sigma\in S_k}
\mathrm{sign}(\sigma)
\tau(a_0\otimes a_{\sigma(1)}\dots\otimes a_{\sigma(k)})
\nonumber \\
&{}&
\times
\mathrm{Tr}(M_0 M_{\sigma(1)}\cdots M_{\sigma(k)}),
\label{maps}
\end{eqnarray}
where $\tau$ is the Hochschild cochain which is a ${\mathfrak s}{\mathfrak p}_{2n}({\mathbb C})$-invariant cocycle in the normalized Hochschild complex 
$
\overline{C}^{2n}({A}_{2n}^{\rm pol}) =
{A}_{2n}^{\rm pol}\otimes 
({A}_{2n}^{\rm pol}/{\mathbb C}\cdot{\bf 1})^{\otimes 2n}
$
\cite{Feigin}.
These maps are compatible with the inclusion 
$\iota_{n'n}: {\rm gl}_n \rightarrow {\rm gl}_{n'}$,
$n < n'$ obtained by embedding an $n \times n$ matrix in
the first rows and columns of an $n'\times n'$ matrix
and completing with zeros. Indeed $\iota_{n' n}$ induces
a restriction map $\iota_{n' n}^*$ on complexes and
one has
$
\varphi^{n}=\iota_{n' n}^*\circ \varphi^{n'}.
$
Recall that $SV =\oplus S^jV$ denote the symmetric algebra
of a $\bf k$-module $V$, and by composing $\varphi^n$ with the dual of the homomorphism of 
${\rm gl}_n(A)$-modules 
one gets
$
S{\rm gl}_n (A)
\longrightarrow {\rm gl}_n (A),\,
z_1\cdots z_k \longmapsto \frac{1}{k!}
\sum_{\sigma\in S_k}z_{\sigma(1)}\cdots z_{\sigma(k)},
$
where the product on the left side is the product in the symmetric algebra of the vector space $S{\rm gl}_n(A)$, while the product on the right side is the associative product of $M_n({\mathbb C})\otimes {A}$.
Extensions of $\varphi^n$ to a chain map for all 
$j\geq 1$ are 
\begin{equation}
\varphi^n_j\,:\, C^\bullet(A) 
\longrightarrow  C^\bullet({\rm gl}_n(A); 
S^j{\rm gl}_n(A)^*).
\label{map}
\end{equation}

Recall definition of the Hochschild cohomology. Let $A$ be an associative algebra over ${\mathbb C}$. The Hochschild cochain complex with coefficients in $A$ is the sequence of vector spaces
$
C^n(A)= {\rm Hom}_{\mathbb C}(A^{\otimes n}, A),\, n=0,1,\ldots
$,
equipped with an operator $d_{\rm Hoch}: C^n(A)\rightarrow C^{n+1}(A)$,
\begin{eqnarray}
(d_{\rm Hoch} f)(a_1,\ldots,a_{n+1}) & = &
a_1 f(a_2,\ldots,a_n)
\nonumber \\
& + & \sum_{j=1}^n (-1)^j f(a_1,\ldots,a_{j-1},
a_j a_{j+1},a_{j+2},\ldots,a_n)
\nonumber \\
& + &  (-1)^{n+1} f(a_1,\ldots,a_n) a_{n+1}. 
\end{eqnarray}
The cohomology of $d_{\rm Hoch}$ in degree $n$ will be denoted ${HH}^n(A)\equiv {HH}^n(A, A)$,
\begin{equation}
{HH}^n(A):=\frac{{\rm Ker}(d_{\rm Hoch}: C^n(A)\longrightarrow C^{n+1}(A))}
{{\rm Im} (d_{\rm Hoch}: \, C^{n-1}(A)\longrightarrow C^n(A))}
\end{equation}
and called {\it the Hochschild cohomology of $A$ with coefficients in $A$}. Suppose that $A$ is a ${\mathbb Z}_2$-graded algebra and $A_p$ be a degree-$p$
component of $A$, such that $A_p\cdot A_q \subset A_{p+q}$.
We say that element $f$ of $C^n(A)$ has an internal degree $p$ if
$f(a_1,\ldots , a_n)\in A_{p+k_1+\cdots+k_n},
$
$a_i\in A_{k_i}$. The vector space $C^n(A)$ is graded by the internal degree, and the total degree of an element has the form
$C^\bullet(A) = \oplus_n C^n(A)$.
The Hochschild complex is graded by the total degree, and the
Hochschild differential can be expressed in the form
\begin{eqnarray}
(d_{\rm Hoch} f)(a_1,\ldots,a_{n+1}) & = &
(-1)^{a\cdot f}\ a_1 f(a_2,\ldots,a_n)
\nonumber \\
& + & \sum_{j=1}^n (-1)^j f(a_1,\ldots,a_{j-1},
a_j a_{j+1},a_{j+2},\ldots,a_n)
\nonumber \\
& + &  (-1)^{n+1} f(a_1,\ldots,a_n) a_{n+1}.
\end{eqnarray}
Let ${\mathfrak A}=(A,Q)$ be a DG-algebra. The degree-1 derivation $Q$ as the map $Q:\, A_p\rightarrow A_{p+1}$
satisfies $Q^2=0$, and is given by
\begin{eqnarray}
&& (Qf)(a_1,\ldots,a_n) =  Q(f(a_1,\ldots,a_n))
\nonumber \\
&& - 
\sum_{j=1}^n (-1)^{v_1+\ldots+v_{j-1}+f+n-1}
f(a_1,\ldots,a_{j-1},Qa_i,a_{j+1},\ldots,a_n).
\label{Qf}
\end{eqnarray}
Each two-cocycle $({Q} f)(a_1,a_2)$ in (\ref{Qf}) defines an infinitesimal deformation of the associative product on $A$. 
Indeed, define a new product by
$
\alpha\circ \beta = \alpha \beta +t ({Q}f)(\alpha, \beta),\, t\in {\mathbb C},
$
then it will be associative to linear order in $t$ iff 
$({Q}f)=0$. Trivial infinitesimal deformations which lead to an isomorphic algebra are classified by two-coboundaries (i.e. two-cocycles of the form $({Q}f)\tau$ for some one-cochain $\tau(\alpha)$). 

Thus ${HH}^2(A)$ classifies nontrivial deformations of the associative algebra structure on $A$. A similar interpretation can be given to the Hochschild cohomology ${HH}^{\bullet}(A)$: it classifies infinitesimal deformations of $A$ in the class of $A_\infty$ algebras (associative algebras being a very special case of $A_\infty$ algebras).

The maps (\ref{map}) induced maps $\varphi^n_{k, j}$ on cohomology:
\begin{eqnarray}
C^\bullet (A) 
& \stackrel{\varphi_j^n}{\longrightarrow} &
C^\bullet({\rm gl}_n(A);\, S^j {\rm gl}_n(A)^*)\,,
\nonumber \\
{HH}^k(A) 
& \stackrel{\varphi^n_{k, j\, (j\geq 1)}}{\longrightarrow}&
H^k({\rm gl}_n(A);\, S^j {\rm gl}_n(A)^*)\,.
\label{filter11}
\end{eqnarray}
\begin{remark}
{\it From {\rm (\ref{commutator})} it follows that 
$
C^\bullet(W_n, {\rm g}{\rm l}_n; {\bf k}) \hookrightarrow C^\bullet(W_n\ltimes {\overline G}\otimes P_n, {\rm g}{\rm l}_n; {\bf k}).
$
Also there is an isomorphism of cohomology rings of reduced relative Weyl algebra and relative cochain complex of infinite dimensional Lie algebra $W_n\ltimes {\overline G}\otimes {P}_n$ over module of any Lie subalgebra ${{\mathfrak g}}\subset ({\rm g}{\rm l}_n\oplus{\overline G})$
{\rm (}see Eq. {\rm (\ref{D1})}{\rm )}:
$[\overline{\alpha}]: \, H^\bullet\widetilde{W}^\bullet
({\rm g}{\rm l}_n\oplus{\overline G}, {{\mathfrak g}})
\stackrel{\simeq}{\longrightarrow} H^\bullet 
(W_n\ltimes {\overline G}\otimes {P}_n, {{\mathfrak g}};
{\bf k})\,.
$
Note that isomorphism $[\overline{\alpha}]$ is not depends form choose of connection $\alpha$. Thus we assume that the modified action of a topological model is  BRST-invariant and the deformed BRST charge is
$
Q = \overline{\partial} + \partial_{\rm deform},
$
where the operator $\partial_{\rm deform}$ is a linear vector field. Its Lie algebra is ${\rm gl}_n$ and the elements of this algebra in coordinates $z_1, \ldots , z_n$ is given in Eq. 
{\rm (\ref{LAlgebra})}. The operator $Q = \overline{\partial} + \partial_{\rm deform}$ satisfies the condition $Q^2 = 0$ and on the vector space $C^\bullet(A)$ there are two comute differentials: $Q$ and $d_{\rm Hoch}$. The Hochschild cohomology of $A$ is defined to be the cohomology of $(-1)^n (\overline{\partial} + \partial_{\rm deform})+d_{\rm Hoch}$.}
\end{remark}
It has been shown that closed topological string states are related to infinitesimal deformations of the open-string theory. The closed string correlators perhaps can be constructed from the open ones using topological string theories as a model. The conjecture is \cite{Kapustin1}: the space of physical closed-string states is isomorphic to the Hochschild cohomology of $(A,Q)$, operator $Q$ of ghost number one (i.e. in DG-algebra). This conjecture has been partially verifyed by means of computation of the Hochschild cohomology of the category of D-branes.

\noindent
One can generalized this statement for the case of perturbative deformations. If in the theory exists a single D-brane then all the information associated with deformations is encoded in an associative algebra $A$ equipped with a differential ${Q} = \overline{\partial} + \partial_{\rm deform}$. Equivalence classes of deformations of these data are described by a Hochschild cohomology of $(A, Q)$, an important geometric invariant of the {\rm (}anti{\rm )}holomorphic structure on $X$.

\section{Deformation pairing}

\subsection{The harmonic structure} 

Suppose that $X$ is a smooth proper variety of dimension $n$ over $\mathbb C$. The vector space structure of the harmonic structure 
$(HT^i(X);\,\, H\Omega_i(X))$ of $X$ is defined as
\begin{equation}
HT^i(X) = \bigoplus_{p+q=i} H^p(X, \Lambda^q TX)\,,
\,\,\,\,
H\Omega_i(X)  = \bigoplus_{q-p=i} H^p(X, \Omega^qX).
\label{harmonic}
\end{equation}
These vector spaces carry the same structures as 
$(HH^i(X),\,\, HH_i(X))$, namely:
$HT^i(X)$ is a ring, with multiplication induced by the exterior product on polyvector fields; 
$H\Omega_i(X)$ is a module over
$HT^i(X)$, via contraction of polyvector fields with forms.
It follows that:  
$HH^2(X)$ contains $H^1(X,TX)$, the space of infinitesimal complex structure deformations; 
$H^0(X,\Lambda^2 TX)$ is a global bivector, giving rise to a noncommutative deformation;
the group $H^2(X,{\cO})$ is a gerbey deformation. 
For a smooth proper variety $X$ $({\rm dim}_{\mathbb C}X = n)$ we will use the following notations: 
\begin{enumerate}[(a)]
\item{} The diagonal embedding
$
\triangle: \, X\hookrightarrow X\times X= X^2,
$
\,
$K_X$ is the canonical bundle of $X$.
\item{}
D$^b({\mathfrak C})$ is equivalent to the full subcategory of 
D$({\mathfrak C})$ consisting of objects $X$ such that $H^n(X)=0$
for $|n|\gg 0$.
\item{}
${\cO}_\triangle = \triangle_*{\cO}_X$ (the structure sheaf of the diagonal in $X\times X$).
The identity functor from D$(X)$ to itself is given by the kernel $\Delta_*{\cO}_X = {\cO}_\Delta$ which is a coherent sheaf. 
We will refer to a sheaf of ${\cO}_X$-modules as an ${\cO}_X$-module. One can take a some of copies ${\cO}_X^{\oplus n} = (\underbrace{{\cO}_X \oplus {\cO}_X \oplus \ldots \oplus {\cO}_X}_n$
to give another ${\cO}_X$-module (the free ${\cO}_X$-module of rank $n$). 
\end{enumerate}
The identification $\bigoplus_{p+q = n}H^p(X, \Lambda^qTX)$ with the group 
${\rm Ext}_{X^2}^n({\cO}_\Delta, {\cO}_\Delta)$ 
is given by the HKR isomorphism between these two 
groups \cite{Hochschild}.
The reader can recognize this result as the space of closed string states in the B-model.

\subsection{Algebra deformations}

The HKR theorem states that for a commutative algebra, $A$, 
$
HH^i(A) \cong  \Lambda^i{\rm Der}(A),
$ 
where Der$(A)$ is the space of derivations of A. There is a map from $A$ to the algebra ${\rm Der}(A)$ of derivations of the algebra defines an exact sequence of algebra homomorphisms
$
0 \rightarrow {\bf k}\rightarrow A \stackrel{\mathfrak m}{\rightarrow}{\rm Der} (A)\rightarrow 0
$
where $\mathfrak m$ is the map $({v} + g)\mapsto [{v} + g,\, 
-]$,\,  ${v}\in {\rm gl}_n,\, g\in {\overline G}$.
The equivalent statement is:
$
HH^i({\rm Spec}(A)) \cong \, H^0({\rm Spec}(A),\Lambda^i T{\rm Spec}(A)),
$
where ${\rm Spec} (A)$ is the set of all proper prime ideals of $A$. Since any variety can be covered by affine patches, one can think of the HKR theorem as a globalization of this result.

Let us consider a geometrical interpretation of the Hochschild cohomology. One can regard an associative algebra $A$ as the algebra of functions on an affine scheme $X={\rm Spec}(A)$.
Then consider $A\otimes A$, its spectrum ${\rm Spec}(A\otimes A)=X^2$, and the diagonal $\triangle\subset X^2$.
One can analyse open-string spectrum of $\triangle$ (i.e. endomorphism algebra); it turns out that the resulting algebra of physical open-string states is precisely the Hochschild
cohomology of $A$. Indeed, the Hochschild cohomology of $A$ is 
$
HH^i(A)= {\rm Ext}_{A\otimes A}^i(A,A);
$
it is the endomorphism algebra of $A$ regarded as an object of the derived category of modules over $A\otimes A$.
\begin{remark}
{\it If $A$ is noncommutative, then $A$ is not a module
over $A\otimes A$, but it is a module over $A\otimes A^{\rm op}$, where $A^{\rm op}$ is the opposite
algebra of $A$. Thus we will have more general case:
$
HH^i(A)= {\rm Ext}_{A\otimes A^{\rm op}}(A,A)
$
for which we need to compute the endomorphisms of $\triangle$ in D$^b(X^2)$. That is, one has to take a projective
resolution of $A$ regarded as a module over $A\otimes A^{\rm op}$, apply to it the operation
${\rm Hom}_{A\otimes A^{\rm op}}(-,A)$, and evaluate the cohomology of the resulting complex of vector spaces.
The main point is that for {\it any algebra} $A$ with a unit there is a canonical resolution of $A$
by free $A\otimes A^{\rm op}$ modules:
$
\cdots \rightarrow A^{\otimes 4}\rightarrow A^{\otimes 3}
\rightarrow A^{\otimes 2}.
$
Each term in this complex is a bimodule over $A$, which is the same as a module over $A\otimes A^{\rm op}$.
Then if we use this resolution to compute ${\rm Ext}^i(A,A)$, we get the Hochschild complex.}
\end{remark}
For an affine space $X$ we have the following result
\begin{eqnarray}
HH^i({\rm Spec (A)})
& \stackrel{\varphi^n_{k, j}}{\longrightarrow} &
H^i({\rm gl}_n(A); S^j{\rm gl}_n(A)^*)\,,
\nonumber \\
HH^i({\rm Spec (A)})
& \stackrel{\cong}{\longrightarrow} &
H^0({\rm Spec} (A), \Lambda^iT{\rm Spec}(A))\,,
\nonumber \\
H^i({\rm gl}_n(A); S^j{\rm gl}_n(A)^*)
& -- &
H^0({\rm Spec} (A), \Lambda^iT{\rm Spec}(A))\,.
\label{D2}
\end{eqnarray}
\\
In the case of the B-model a bulk-boundary OPE could be a pairing
\begin{equation}
H^0(X, \Lambda^iTX) \times {\rm Ext}_X^n({\cE}, {\cF})\rightarrow
H^i({\rm gl}_n(A); S^j{\rm gl}_n(A)^*) \times {\rm Ext}_X^n({\cE}, {\cF})
\rightarrow {\rm Ext}_X^{n+i}({\cE}, {\cF})
\end{equation}
It is natural to conjecture that such a mathematical pairing realizes some bulk-boundary OPE above under deformation of the BRST operator. Perhaps any mathematical (or geometrical) deformations of a sheaf match physical deformations of the corresponding branes in the sigma model.
In the special case that the sheaves are bundles on $X$,
so that the Ext groups reduce to sheaf cohomology on $X$ represented by differential forms, the mathematical pairing reduces to a wedge product of differential forms, just like the Yoneda pairing in such circumstances and the bulk-bulk OPE in all circumstances. This bulk-boundary pairing is defined as follows:
\begin{enumerate}[{-}]
\item{} First, we must define a map 
$
HH^i(X)\rightarrow H^i({\rm gl}_n(A); S^j{\rm gl}_n(A)^*) \rightarrow
\mbox{Ext}_X^{i}( {{\cE}}, {{\cE}})
$
that maps bulk states (in the presence of deformations) to states defined on the boundary.
\item{} Then, we use the Yoneda pairing 
$
\mbox{Ext}_X^{i}\left( {{\cE}}, {{\cE}} \right) \times
\mbox{Ext}^n_X\left( {{\cE}}, {{\cF}} \right) \rightarrow
\mbox{Ext}^{n+i}_X\left( {{\cE}}, {{\cF}} \right).
$
\end{enumerate}
For the first map we need identify $HH^i(X)$ (and therefore 
$H^i({\rm gl}_n(A); S^j{\rm gl}_n(A)^*)$) with the group 
${\rm Ext}_{X^2}^i({\cO}_\Delta, {\cO}_\Delta)$).
Then we need define a pairing on
$H\Omega_i(X)$ which is a modification of the usual pairing of forms given by cup product and integration on $X$.  
Given the morphism above, we can define
the desired bulk-boundary map. 
Note that a Fourier-Mukai transform with kernel
${{\cO}}_{\Delta}$ maps ${{\cE}}$ to ${{\cE}}$, and with kernel
${{\cO}}_{\Delta}[n]$ maps ${{\cE}}$ to ${{\cE}}[n]$.
A bulk state identified via the morphism above as an element of 
$
\mbox{Ext}^n_{X^2} \left( {{\cO}}_{\Delta}, 
{{\cO}}_{\Delta} \right)
= \mbox{Hom}_{X^2}\left( {{\cO}}_{\Delta},
{{\cO}}_{\Delta}[n] \right)
$
is a map ${\mu}: {{\cO}}_{\Delta} \rightarrow {{\cO}}_{\Delta}[n]$.
Finally a map between the kernels of two Fourier-Mukai transforms
defines a map ${{\cE}} \rightarrow {{\cE}}[i]$ between the images
of a given object ${{\cE}}$, and the bulk state ${\mu}$ defines an element of
$
\mbox{Hom}_X\left( {{\cE}}, {{\cE}}[n] \right) = 
\mbox{Ext}^n_X\left( {{\cE}}, {{\cE}} \right).
$
Given an element of $H^i({\rm gl}_n(A); S^j{\rm gl}_n(A)^*)$ one can define an element of $\mbox{Ext}^{p+i}_X\left( {{\cE}},{{\cE}} \right)$.

\subsection*{Acknowledgments}

A. A. Bytsenko would like to thank the Conselho Nacional de Desenvolvimento Cient\'ifico e Tecnol\'ogico (CNPq) for support.

\end{document}